# Predictability of a system with transitional chaos

Marek Berezowski

Cracow University of Technology, Faculty of Chemical Engineering, Cracow, Poland
(E-mail: marek.berezowski@pk.edu.pl)

**Abstract.** The paper is focused on the discussion of the phenomenon of transitional chaos in dynamic autonomous and non-autonomous systems. This phenomenon involves the disappearance of chaotic oscillations in specific time periods and the system becoming predictable. Variable dynamics of the system may be used to control the process. The discussed example concerns a model of a chemical reactor.

**Keywords:** Chaos, Chemical Reactor, Oscillations, Time series, Lyapunov time.

## 1 Introduction

Unpredictability is characteristic for a chaos. It means, that even the smallest change in initial conditions causes that we are unable to predict what will happen in the future. The greater change, the shorter the Lyapunov time (predictability). However, there are completely opposite cases, that is: we are unable to predict what happen in a moment, but we can say what happen in farther future with 100% accuracy.

It is probable that this feature is embedded in the Universe and nature. Due to a chaotic nature of the Universe, we are not even capable of predicting the weather prevailing on the Earth, or the movement of the galaxies, yet we forecast that in the far future the Universe will shirk again back to the unique point and shall be predictable. It is similar with nature. Until an entity lives, it is difficult to predicts its behavior, but when it dies, its condition is explicitly determined within the framework of time. Thus, we may say what will happen to an entity/ individual in the far future, yet, we may not foresee what will happen to them in a moment of time. It occurs in systems with variable dynamics. There may be autonomous or non-autonomous systems. Under certain conditions the phenomenon of variable dynamics may be used to predict chaotic changes and to control a given process.

## 2 Autonomous system

An example of the autonomous system can be two-dimensional discrete model:

$$x_{k+1} = 4x_k\left(1 - x_k - \frac{0.001}{y_k}\right)$$

$$y_{k+1} = \frac{y_k}{2.15}(1 - x_k^2). \qquad (1)$$

See Fig. 1.

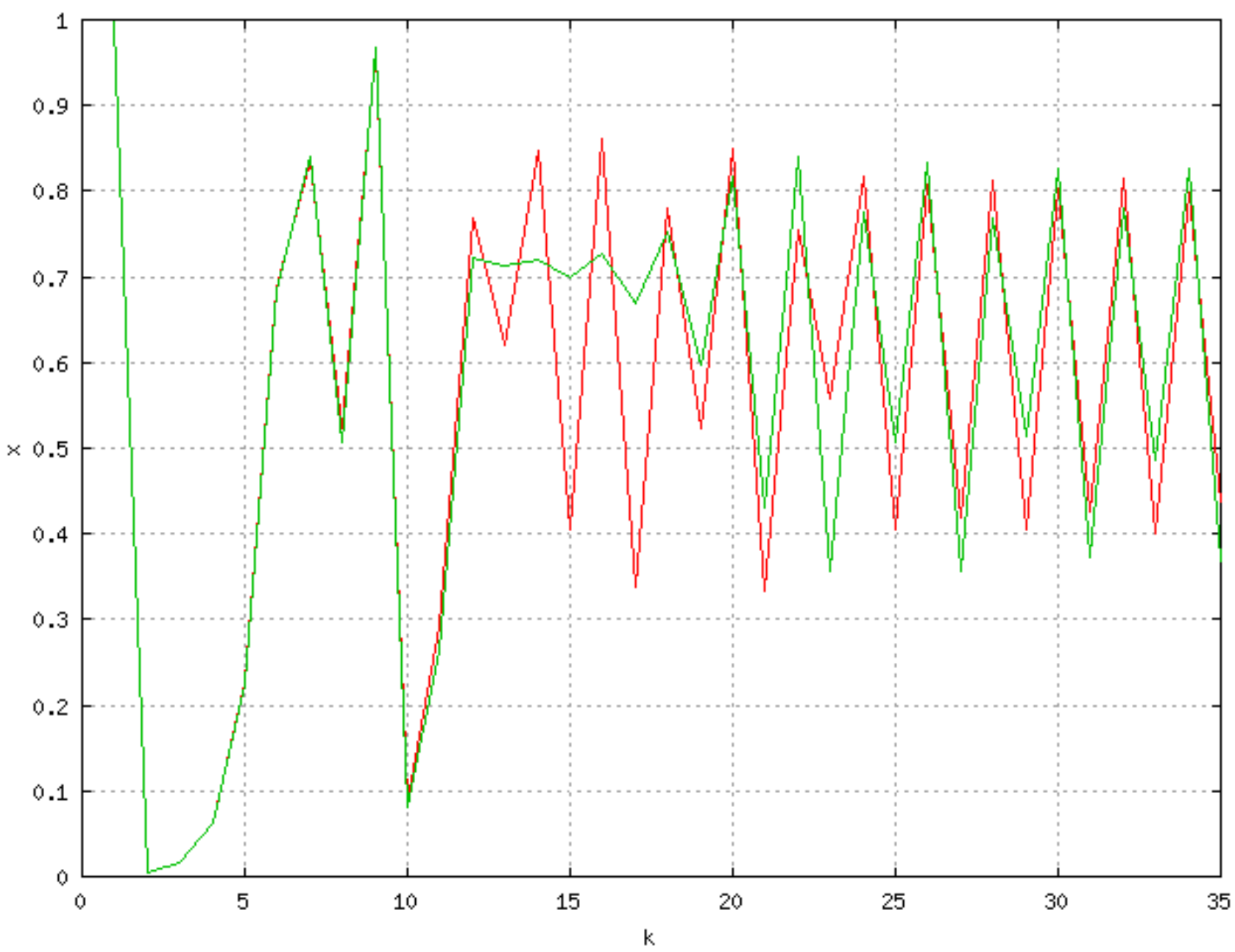

Fig. 1. Two time series of the autonomous system (1). Green line – first plot, red line – second plot

Two trajectories start at slightly different initial conditions, and, for k<12 they are practically convergent. For k>12; however, the trajectories become distant from each other. This is a typical feature of chaos. Yet, for k>26 both trajectories are again approaching each other and from that moment the system generates stable (predictable) four-period oscillations.

A similar phenomenon may occur in different types of industrial systems and equipment. A good example of this is a chemical reactor in which the so called catalyst deactivation takes place. Due to such deactivation the reactor slowly halts, and after some time, stops working at all. If such process is considered in a mathematical model of the reactor, and described by means of additional differential equations, we are dealing with an autonomous model. If, however, the deactivation process is explicitly dependent on time, we are dealing with a non-autonomous model.

## 3 Non- autonomous system

An example of a non-autonomous model is a chemical reactor with mass recycle, in which the following reaction kinetics variable in time was assumed:

$$\begin{aligned}
&\frac{\partial a}{\partial \tau}+\frac{\partial \alpha}{\partial \xi}=(1-f)\phi(\alpha,\Theta) \\
&\frac{\partial \Theta}{\partial \tau}+\frac{\partial \Theta}{\partial \xi}=(1-f)\phi(\alpha,\Theta)+(1-f)\delta(\Theta_H-\Theta) \\
&\alpha(\tau,0)=f\alpha(\tau,1);\Theta(\tau,0)=f\Theta(\tau,1) \\
&\phi(\alpha,\Theta)=Da(1-\alpha)^n e^{\gamma\frac{\beta\Theta}{1+\beta\Theta}}\left(e^{-\omega_1 k}-e^{-\omega_2 k}\right)
\end{aligned} \qquad (2)$$

where: *f=0.5,* $\delta=3$, $\Theta_H=-0.025$, *Da=0.15, n=1.5,* $\gamma=15$, $\beta=2$, $\omega_1=0.00025$, $\omega_1=0.0075$ and $k$ is a disctrete time. See Fig. 2.

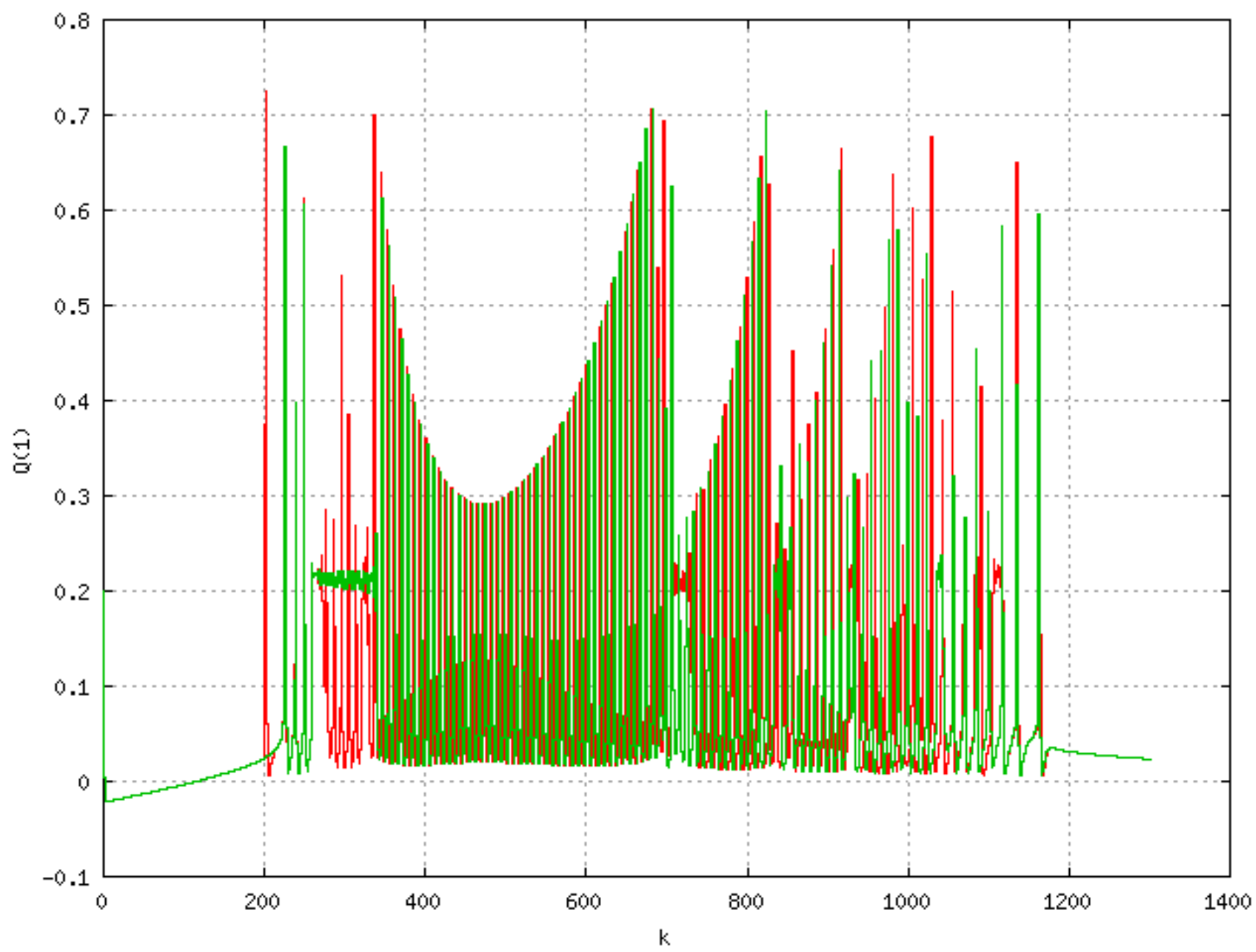

Fig. 2. Two time series of the chemical reactor model (2). Perturbation for $k=200$. Green line – first plot, red line – second plot

Likewise, two trajectories start at only slightly different initial conditions, and, for $k<250$ they are practically convergent. But, for $k>250$ they are becoming divergent, which is a typical feature of chaos. For the range of $340<k<680$ the process becomes completely predictable, see Fig. 3.

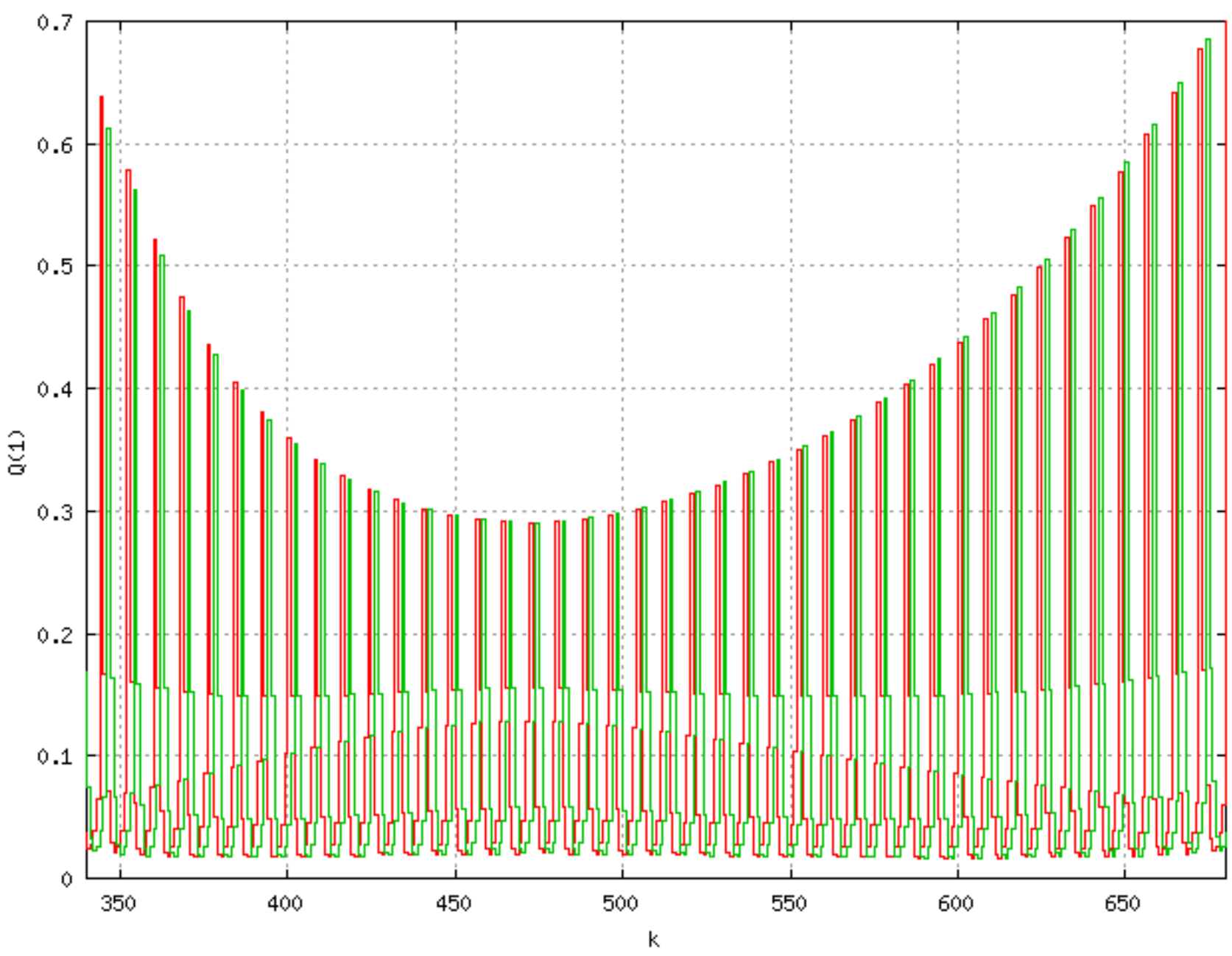

Fig. 3. Part of Fig. 2.

For $680<k<1180$ chaos reappears and the resultant unpredictability (chaotic crisis). For $k>1180$ the system enters the steady state and again becomes predictable. In Fig.4 Poincaré section is shown for the function in Fig. 2. This is Henon characteristic attractor, designated for a reactor with recycle- see [1-12 ].

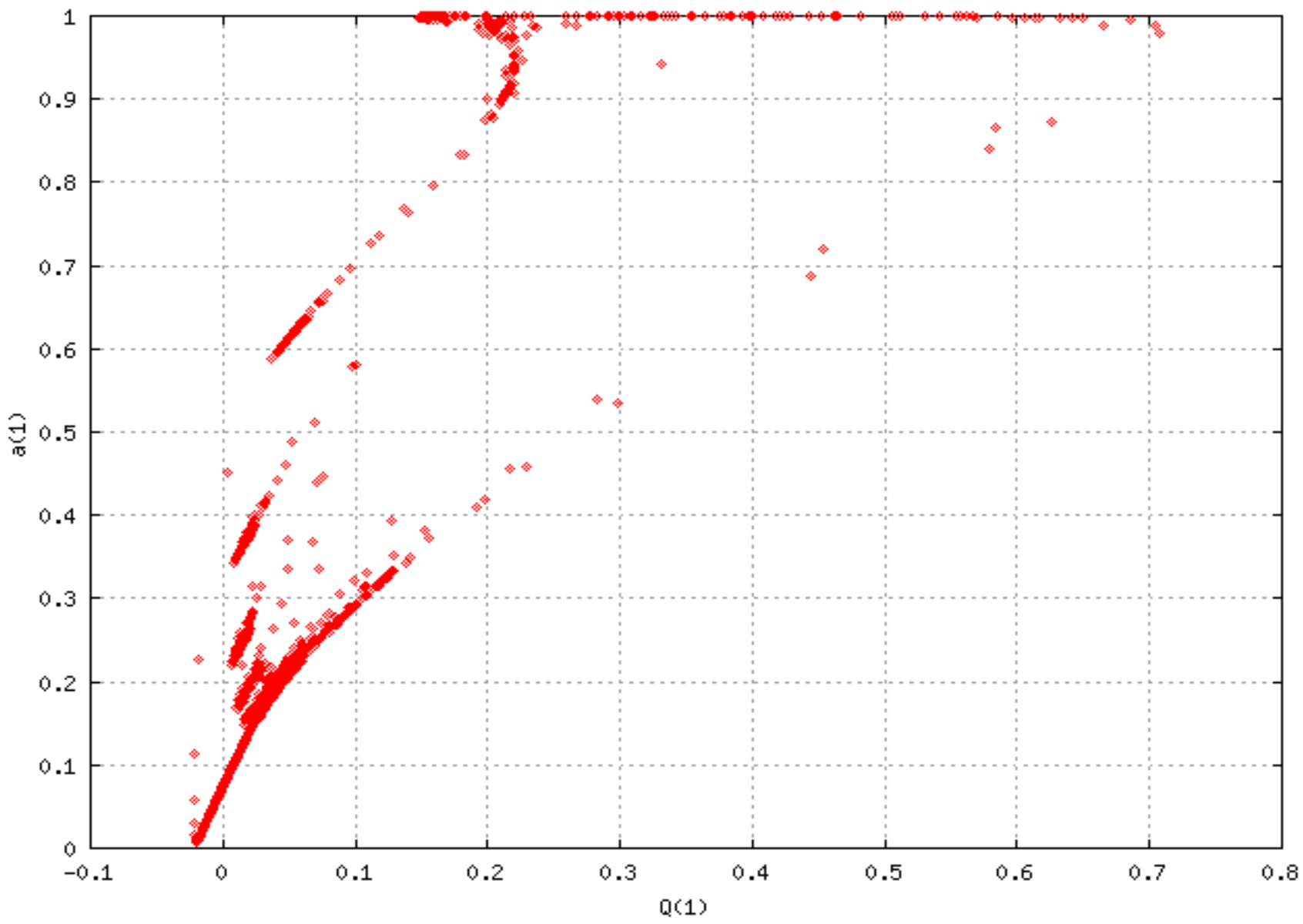

Fig. 4. Poincaré set of the chemical reactor model (2)

It is noteworthy that for perturbation in time k<200 does not evoke changes in the further course of the process. See Fig. 5.

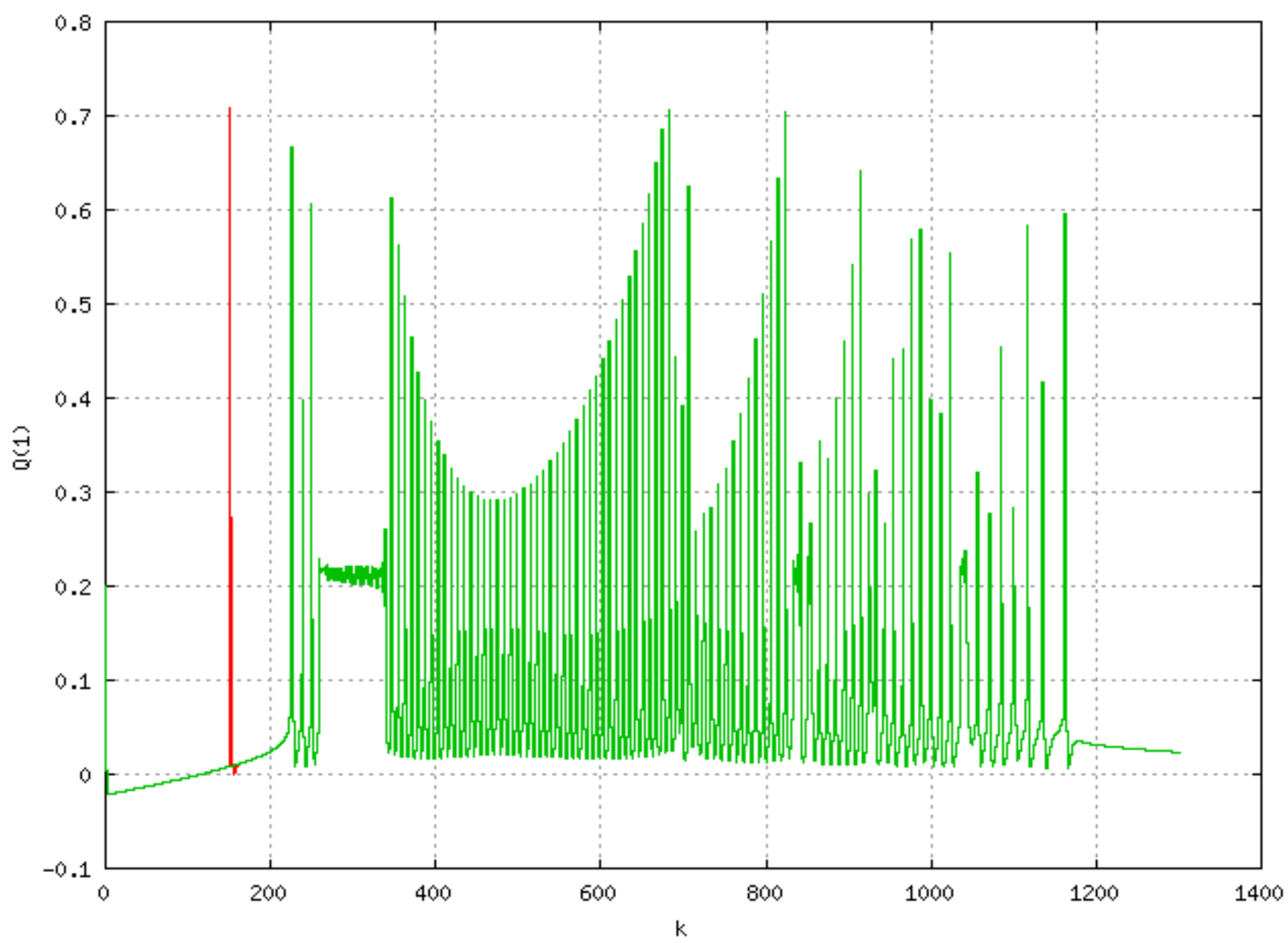

Fig. 5. Two time series of the chemical reactor model (2). Perturbation for k=150. Continuous line – first plot, dashed line – second plot

The phenomenon of variable dynamics may be used to predict chaos and, in consequence, enable process control. Let us assume the following kinetics form of the reaction:

$$\phi(\alpha,\Theta) = Da(1-\alpha)^n e^{\gamma\frac{\beta\Theta}{1+\beta\Theta}} e^{-\frac{\omega}{k}} \qquad (3)$$

If, under normal process conditions, i.e. for $\omega = 0$, the reactor works in a chaotic manner, then, assuming $\omega \neq 0$, the value of function $\phi$ is close to zero for small values of time k. This means that during the first time interval, the reactor works in a predictable manner. During the start-up the values of the reactor state variables (temperature and concentration) are the same for both trajectories, irrespective of the difference in their initial conditions. As shown in Fig. 6, for $\omega = 20$ even a significant perturbation in time k=10 does not evoke changes in the further course of the process.

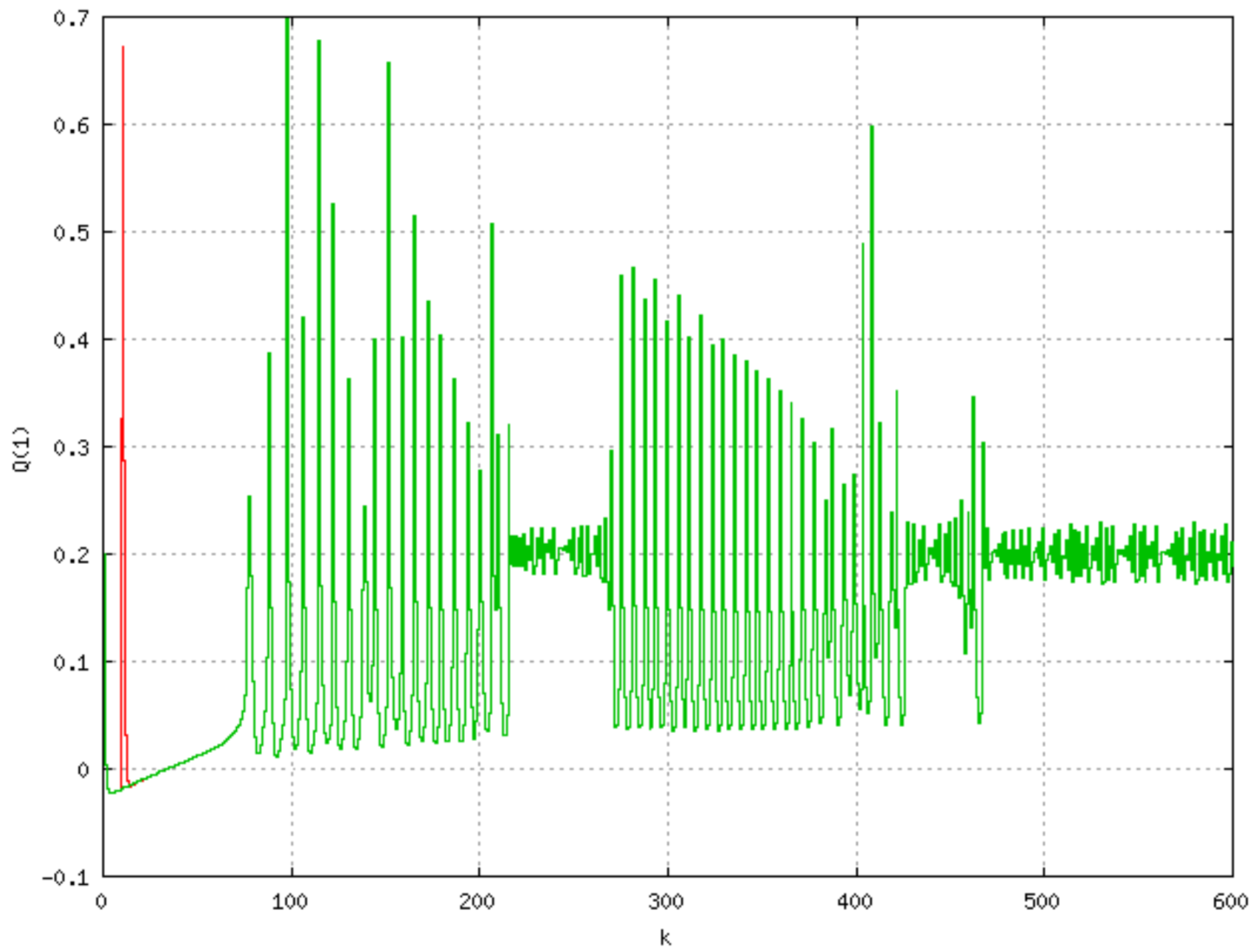

Fig. 6. Two time series of the chemical reaktor model (3). Perturbation for k=10. Continuous line – first plot, dashed line – second plot

This means that such chaotic process does not react to changes in its initial conditions, thus, it is predictable.

The occurrence of a disturbance in time k>10 does not lead to changes in the beginning for a significantly long time period. However, the changes appear in longer time perspective, as shown in Fig. 7.

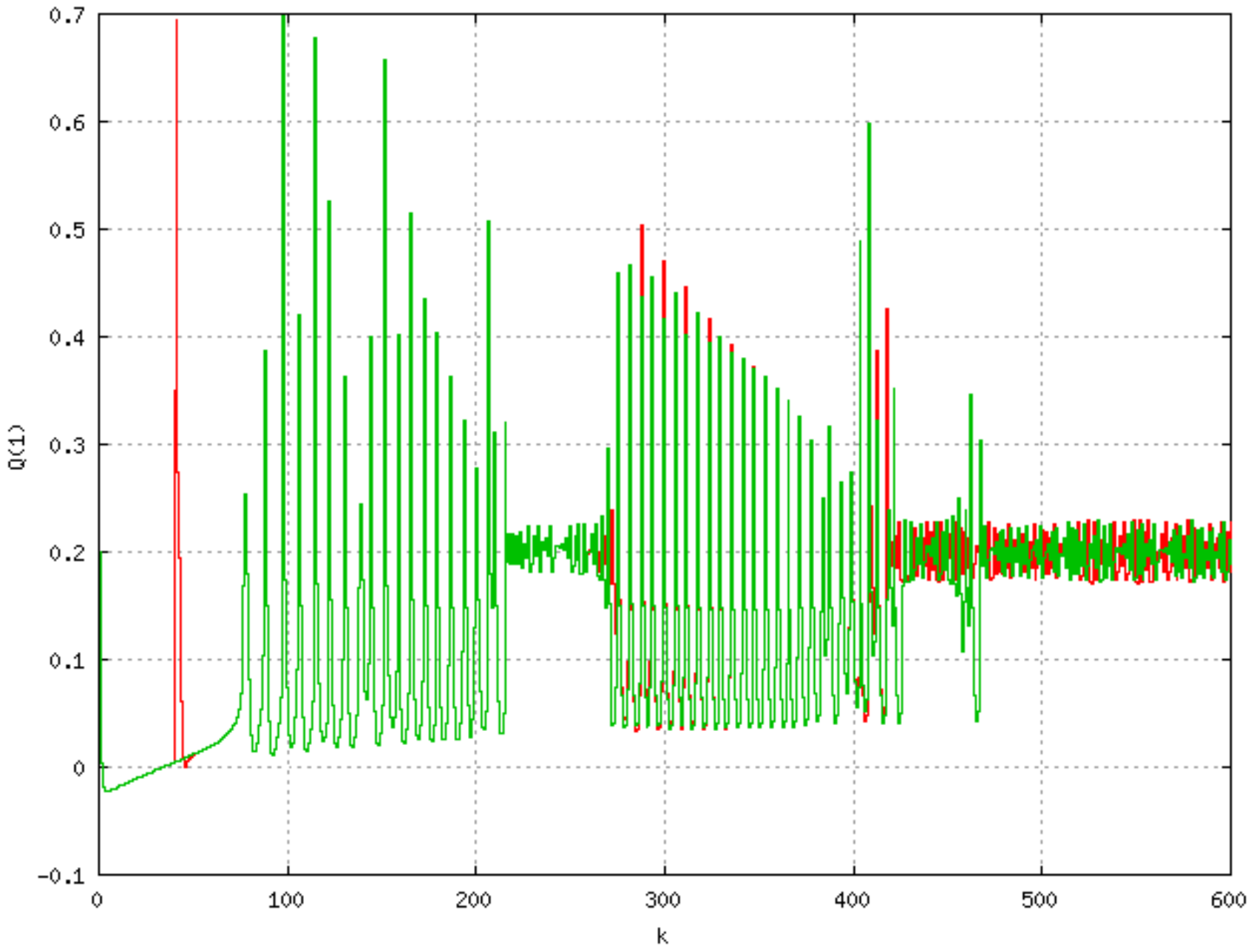

Fig. 7. Two time series of the chemical reactor model (3). Perturbation for k=40. Continuous line – first plot, dashed line – second plot

Accordingly, the excitation of non-chaotic operation of the reactor at start-up may guarantee the predictability of the process in its further course. This may be useful in circumstances when the reactor is required to work chaotically, yet at the same time, under control.

## Conclusions

The scope of the paper is the analysis of the phenomenon of transitory chaos in dynamic autonomous and non-autonomous systems. In the autonomous system chaos occur in a transitory state and then disappears in the steady state. Accordingly, the state of the system is unpredictable at the beginning of the observations, but predictable in the longer time period. Likewise, in a non-autonomous model explicitly dependent on time. Such case was considered on the example of a model of a chemical reactor. The phenomenon of variable dynamics may be used to predict chaotic behaviour of systems, which, in turn, enables their control.

The system (2) remembers the initial conditions for $k < 200$. The system (3) remembers the initial conditions for $k < 10$. We can say that these are special times of Lapunov.